\documentclass[conference]{IEEEtran}
\IEEEoverridecommandlockouts
\usepackage{cite}
\usepackage{amsmath,amssymb,amsfonts}
\usepackage{algorithmic}
\usepackage{graphicx}
\usepackage{textcomp}
\usepackage{xcolor}
\usepackage{booktabs}
\usepackage{adjustbox}
\usepackage{listings}
\usepackage{url}
\usepackage{flushend} 
\usepackage{comment}
\usepackage{tikz}
\usepackage{todonotes}
\usepackage{enumitem}

\usetikzlibrary{positioning,arrows.meta, shapes, fit, backgrounds, calc}

\usepackage[switch, mathlines]{lineno}

\usepackage{xcolor} 
\usepackage{soul} 
\usepackage{amssymb} 

\newcommand{\chan}[1]{{\color{black} #1}} 
\newboolean{showcomments}
\setboolean{showcomments}{false} 
\ifthenelse{\boolean{showcomments}}
 { \newcommand{\mynote}[2]{
      \fbox{\bfseries\sffamily\scriptsize#1}
        {\small$\blacktriangleright${{{#2}\bf }}$\blacktriangleleft$}}}
        { \newcommand{\mynote}[2]{}}

\newcommand{\as}[1]{\mynote{Asma}{\hl{#1}}}

\def\BibTeX{{\rm B\kern-.05em{\sc i\kern-.025em b}\kern-.08em
    T\kern-.1667em\lower.7ex\hbox{E}\kern-.125emX}}
\begin{document}



\title{Round-Trip Mutation Testing: Translating Code to Natural Language Intent and back}

\newcommand{\inlineheadingbf}[1]{\medskip\noindent{\bfseries #1.}}
\newcommand{\inlineheadingit}[1]{\medskip\noindent{\em #1.}}

\definecolor{pykeyword}{RGB}{0,0,180}
\definecolor{pycomment}{RGB}{0,120,0}
\definecolor{pystring}{RGB}{163,21,21}
\definecolor{pybg}{RGB}{248,248,248}

\lstdefinestyle{pythonstyle}{
    language=Python,
    backgroundcolor=\color{pybg},
    basicstyle=\ttfamily\footnotesize,
    keywordstyle=\color{pykeyword}\bfseries,
    commentstyle=\color{pycomment}\itshape,
    stringstyle=\color{pystring},
    showstringspaces=false,
}

\lstnewenvironment{python}[1][]
{
    \lstset{style=pythonstyle,#1}
}
{}

\author{

\IEEEauthorblockN{
Asma Hamidi$^1$, 
Cedric Richter$^1$, 
Ahmed Khanfir$^{2,1}$, 
Mike Papadakis$^1$
}

\IEEEauthorblockA{
\{asma.hamidi, cedric.richter, michail.papadakis\}@uni.lu \\
ahmed.khanfir@ensi-uma.tn
}

\IEEEauthorblockA{
$^1$ \textit{SnT, University of Luxembourg}, Luxembourg \\
$^2$ \textit{RIADI, ENSI, University of Manouba}, Tunisia
}

}

\maketitle

\begin{abstract}
This paper presents 
Round-Trip Mutation Testing (RTM), a novel 
approach that generates mutants from LLM mistranslations between a program code and its intent. 
Leveraging the generative capability of LLMs from and to programming and natural language, and given an input program, our approach predicts its intent, that is used to generate programs, which when different from the original one, constitute the output mutants. 
The approach produces additionally mutants, stemming 
from artificially provoked mistranslations, by mutating the intent prior to the final programs (mutants) generation. 
Originating from the propagation of small changes in the intent to the code, our intuition is that these programs would present subtle semantic differences from the original one, simulating likely-to-occur faults that could result from specification misunderstandings, and enabling mutation testing. 
To evaluate RTM, we run it on 40 real buggy methods and evaluate its effectiveness and cost-efficiency in guiding testing towards detecting the bugs.
Our results demonstrate the potential of round-trip mutation testing to produce syntactically more diverse mutants, potentially exposing faults that traditional mutation operators fail to reveal. More interestingly, RTM outperforms traditional pattern-based mutation in producing smaller and stronger test-suites, detecting on average over \chan{4} and \chan{1.7} times more faults when selecting only \chan{4 and 30} tests respectively. 

\end{abstract}


\section{Introduction}

Several studies have proven that mutation testing is efficient in guiding testing towards higher fault-detection capabilities~\cite{papadakis2015metallaxis,PapadakisK00TH19}. 
It operates typically by generating different variants, so-called {\em mutants}, of the program under test by introducing syntactic alterations of its source code. Mutants are evaluated with respect to the current test suite, and a mutant is detected (or killed) if the test suite behaves differently on the mutant than on the original program.
Undetected (or survived) mutants indicate potential untested corner-cases which the developer can cover by adding new targeted tests.

Throughout the last decades, mutation testing has been an important research topic in software engineering~\cite{0020331, PapadakisK00TH19,OffuttLRUZ96, MarcozziBKPPC18, KintisPPVMT18}, leading to the proposition of numerous generation techniques~\cite{mujava,major,pitest} aimed primarily to reduce the cost of mutation analysis by minimizing redundancy while increasing coverage, ensuring diversity among generated mutants, and improving their realism and practical relevance, e.g.  
by simulating real-faults and resembling developer mistakes~\cite{DBLP:conf/icst/DegiovanniP22,khanfir2023efficient, SemSeed, khanfir2020ibir}.

\begin{figure}
    \centering
    \includegraphics[width=\columnwidth]{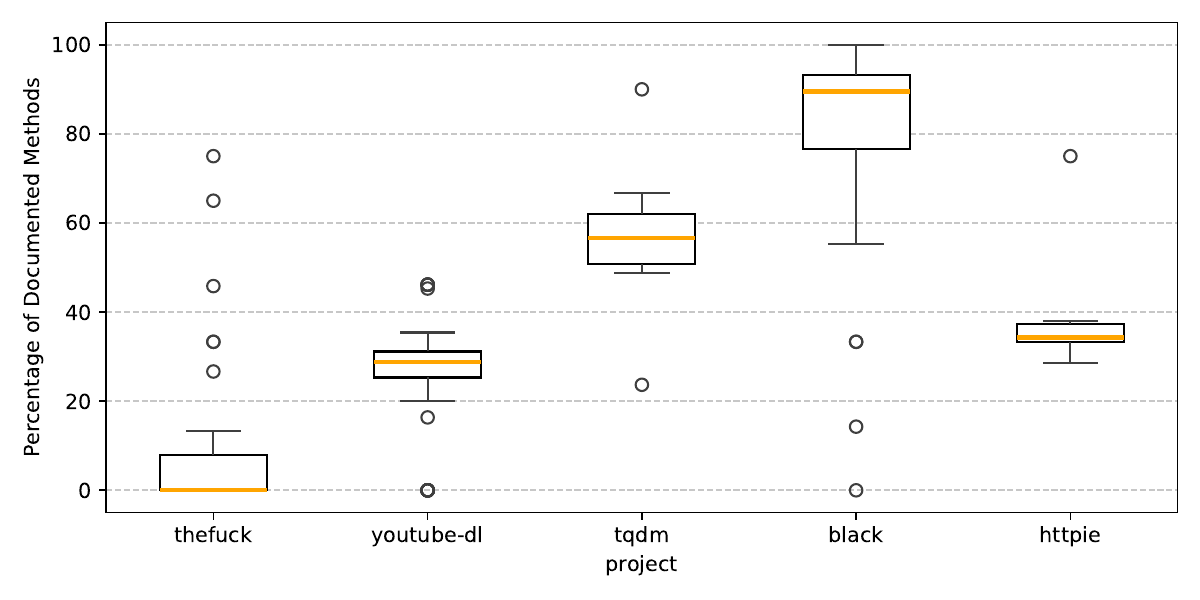}
    \vspace{-2.5em}
    \caption{Distribution of documented methods across the 5 projects}
    \label{fig:documented_methods}
    \vspace{-1em}
\end{figure}

While existing approaches operate at source code level, the emergence of Large Language Models (LLMs) led to the introduction of Intent-based mutation~\cite{hamidi2025intentbasedmutation}, which applies the mutation to the underlying program intent. 
Not starting from the existing code and not relying on predefined mutation operators, enable the 
generation of completely new implementations, forming syntactically different and diverse mutants. 
To this end, intent-based mutation testing mutates the program intent, e.g. represented by the code documentation, and asks an LLM to generate code that corresponds to the mutated intents. 
By mutating the program’s described semantics (intent), the approach is more likely to produce semantically dissimilar mutants from the original program that simulate faults arising from a developer's misunderstanding of the intent of the program.
In practice, developer-provided descriptions (i.e. as method docstrings) are often scarce, limiting the applicability of intent-based mutation testing. In fact, only 34.2\% of the methods in our evaluated real-world projects contain a docstring, as illustrated in Figure~\ref{fig:documented_methods}. 



%

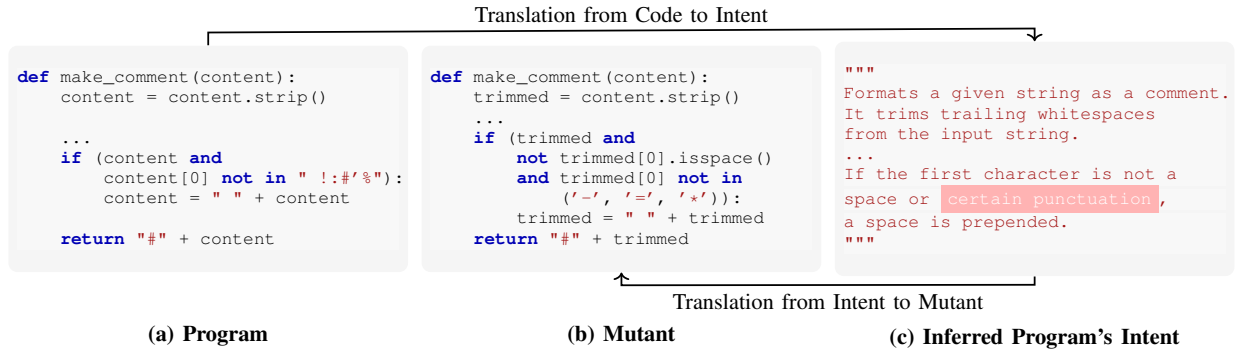
\begin{figure*}
\vspace{-3em}
    \centering
    \begin{adjustbox}{max width=0.9\textwidth}
        \begin{tikzpicture}

\node (code) [anchor=north west, fill=gray!8!white, rounded corners, inner sep=4pt] at (0, 0) {%
\begin{minipage}{0.33\linewidth}
\begin{python}[numbers=none]
def make_comment(content):
    content = content.strip()
    
    ...
    if (content and 
        content[0] not in " !:#'%"):
        content = " " + content
        
    return "#" + content
\end{python}%
\end{minipage}
};
\node[below=0.7cm of code]{\textbf{(a) Program}};

\node[right=0.2cm of code] (mutant) [fill=gray!8!white, rounded corners, inner sep=4pt]{%
\begin{minipage}{0.33\linewidth}
\begin{python}[numbers=none]
def make_comment(content):
    trimmed = content.strip()
    ...
    if (trimmed and 
        not trimmed[0].isspace()
        and trimmed[0] not in 
            ('-', '=', '*')):
        trimmed = " " + trimmed
    return "#" + trimmed
\end{python}%
\end{minipage}
};
\node[below=0.7cm of mutant] (mutant-label) {\textbf{(b) Mutant}};

\node[right=0.2cm of mutant] (intent) [fill=gray!8!white, rounded corners, inner sep=4pt]{%
\begin{minipage}{0.33\linewidth}
\begin{python}[numbers=none, escapechar=!]
"""
Formats a given string as a comment.
It trims trailing whitespaces 
from the input string.
...
If the first character is not a 
space or !\colorbox{red!30!white}{\color{white}certain punctuation}!, 
a space is prepended.
"""
\end{python}%
\end{minipage}
};
\node[below=0.7cm of intent] (intent-label){\textbf{(c) Inferred Program's Intent}};

\draw[->, thick] (code.north) -- ($(code.north) + (0, 0.2cm)$) --  node[midway, above]{Translation from Code to Intent} ($(intent.north) + (0, 0.2cm)$) -- (intent);
\draw[->, thick] (intent.south) -- ($(intent.south) + (0, -0.13cm)$) --  node[midway, below]{Translation from Intent to Mutant} ($(mutant.south) + (0, -0.2cm)$) -- (mutant);

\end{tikzpicture}
    \end{adjustbox}
    \vspace{-1em}
    \caption{\textbf{Round-trip mutation via Large Language Models.}} 
    \label{fig:example}
    \vspace{-1em}
\end{figure*}

To address this limitation, we introduce {\em round-trip mutation testing} (RTM), a novel approach that mutates a program by performing a round-trip translation between code and a natural language description of its intent. 
Our approach relies on the observation that, while large language models (LLMs) are effective in generating code and code-intent, they may introduce noise and imperfections in the translation steps, thereby, deviating the semantics of the output programs from the original one. As a result, 
we obtain mutants 
representing possible 
misunderstandings of the program's behavior. Unlike prior works that use round-trip translation for LLM evaluation~\cite{allamanis2024unsupervised} and program repair~\cite{ruiz2024novel}, our approach leverages this process explicitly to generate mutants.
Figure \ref{fig:example} illustrates an overview of round-trip mutation testing on an example program, where the subfigures \textit{(a)}, \textit{(c)} and \textit{(b)} are respectively: the original code, its inferred intent and the resulting mutant.
As can be seen, the translation from \textit{(a)} to \textit{(c)} is relatively accurate but misses the type of punctuation to check for, leading to the generation of  
code (mutant: from \textit{(c)} to \textit{(b)}) that is semantically close to the original implementation, but checks for different punctuation as in the original code. 

%
%
We implemented four variants of our approach --mutating or not the program intent and asking for a precise or broad intent-- and evaluated its effectiveness and cost-efficiency in finding 40 faults from the BugsInPy dataset~\cite{widyasari2020bugsinpy}, i.e. guiding the selection of fault-revealing tests.  
To do so, we generate mutants on the buggy version of the program and select tests to kill them -- simulating an actual testing scenario, where the bug is neither detected nor fixed. 
The results show that our approach 
can generate valid mutants that are syntactically more diverse than pattern-based mutation testing.
More importantly, we observe that round-trip mutation outperforms traditional mutation testing, in selecting small test suites of higher fault detection capabilities. For example, when we restrict the selection to \chan{4} tests, \chan{RTM} yields 
test suites achieving on average \chan{$\approx$4.3} times higher fault-detection rates than the test suites selected by traditional mutants. 
When more tests are selected, this difference remains noticeable but gradually decreases, i.e., to \chan{$\approx$1.7} times with \chan{30 tests} until it becomes small after the selection of \chan{34 tests}, where \chan{RTM} achieves its maximum avg fault detection of \chan{44.6\%}, which is slightly higher than the \chan{44\%} of traditional mutants.
This endorses the potential of RTM in identifying fault-revealing tests earlier, making it suitable for the efficient construction of small test suites with large fault detection capabilities, while leaving room for improving its effectiveness in larger test suite scenarios. 



In general, our contributions can be summarized as follows:
\begin{itemize}
    \item We propose round-trip mutation testing, a novel approach that produces mutants from LLM mistranslations between natural and programming language, with a generalization of intent-based mutation to real-world scenarios, i.e. enabling it on
    non-documented and description-free code. 
     \item We show that round-trip mutation can generate valid mutants that are syntactically distinct from traditional mutants.
    \item We provide empirical insights on the potential advantage of guiding test selection by round-trip mutation over pattern-based mutation, in achieving small test suites of higher fault detection capability.
 
\end{itemize}

\section{Background}
\label{sec:background}
Mutation testing~\cite{PapadakisK00TH19} is a test-adequacy criterion that evaluates test suite effectiveness based on its capacity to detect mutants. 
These mutants are typically obtained by 
modifying the target program code, 
using predefined and tuned mutation operators\cite{0020331, OffuttLRUZ96,PapadakisK00TH19} which introduce small syntactic 
changes, e.g. replacing an arithmetic operator \texttt{+} with a \texttt{-}. 
A mutant is said to be {\em killed} by a test suite if its execution result on it differs from that on the original program, otherwise, it is said to be {\em survived}~\cite{MarcozziBKPPC18, KintisPM15}.
Test suite thoroughness can then be approximated via its ability to detect ({\em kill}) mutants, where a higher mutant-killing ratio ({\em mutation score}) indicates a higher fault detection capability~\cite{ojdanic2023syntactic}.  
Survived mutants can be used as test-writing or generation target (writing tests to kill mutants), thereby, improving test suite thoroughness~\cite{ojdanic2023comparing}.
Mutation testing introduces also {\em duplicate} and 
semantically {\em equivalent} mutants to the original program. These form a major concern in mutation, falsifying mutation scores and causing unnecessary computational (i.e. test execution) and manual (i.e. analysis) overhead. Hence, the goal for effective mutation testing is to generate distinct {\em killable} mutants that can be used to construct adequate test suites.


\section{Approach}


\subsection{Round-Trip Mutation}
Given two domains $\mathbb{PL}$ (programs) and $\mathbb{NL}$ (natural language descriptions), round-trip mutation generates mutants by translating from $\mathbb{PL}$ to $\mathbb{NL}$ and back. To this end, round-trip mutation employs two models $M: \mathbb{PL} \rightarrow \mathbb{NL} $ and $M^{-1}: \mathbb{NL} \rightarrow \mathbb{PL}$ that ensure the transformation between the two domains. Given a program $P \in \mathbb{PL}$, 
a round-trip mutant $P_m$ is thus defined as:
\begin{equation}
\label{eq:directgen}
    P_m = M^{-1}(M(P)),
\end{equation}
where $P_m$ is syntactically different from $P$. As can be seen in Figure \ref{fig:example}, round-trip mutation testing with an {\em imperfect} translator, i.e. an LLM, can already yield valid mutants of the original program.

In addition to mutants resulting from $M^{-1}$ and $M$ imperfect translations, our approach generates mutants by propagating intent-mutations into code. 
To do so, it applies a transformation $\mu: \mathbb{NL} \rightarrow \mathbb{NL}$ on the output of $M$ (the program intent) before proceeding with the $M^{-1}$ transformation (back to code). 

More generally, for a given input program $P$, our approach generates mutants $P_{\mu}$ as follows:
\begin{equation}
    P_{\mu} = M^{-1}(\mu(M(P)))
\end{equation}
In the simplest case, $\mu$ is the identity function and thus $P_{\mu} = P_m$ (output of Equation~\ref{eq:directgen}). 
%
In our RTM first implementation, we use the same LLM for all transformations ($M$, $M^{-1}$ and $\mu$) and follow Hamidi et al.~\cite{hamidi2025intentbasedmutation} approach in performing intent-mutation ($\mu$), asking the LLM to propose word replacements that modify the intent’s semantics while preserving grammatical correctness and semantic coherence.

\subsection{Prompt Design} 
\begin{figure}
\vspace{-3em}
    \centering
    \begin{adjustbox}{max width=0.9\textwidth}
        \begin{tikzpicture}

\node (broad) [anchor=north west, fill=gray!8!white, rounded corners, inner sep=6pt] at (0,0) {%
\begin{minipage}{0.9\linewidth}
\footnotesize

You are given a Python function \colorbox{gray!15}{\texttt{{function\_name}}}:

{\texttt{```python\\[2mm]
\colorbox{gray!15}{\texttt{{function\_code}}}
\\[2mm]
```
}
 }

in a few sentences, generate a high-level description that explains the overall purpose or intent of the function, focusing on the general goal rather than implementation details.

\end{minipage}
};
\node[below=0.05cm of broad]{\footnotesize (a) Broad};

\node (precise) [anchor=north west, fill=gray!8!white, rounded corners, inner sep=6pt] at ([yshift=-0.5cm]broad.south west) {%
\begin{minipage}{0.9\linewidth}
\footnotesize

You are given a Python function \colorbox{gray!15}{\texttt{{function\_name}}}:

{\texttt{```python\\[2mm]
\colorbox{gray!15}{\texttt{{function\_code}}}
\\[2mm]
```
}
 }

in a few sentences, generate a precise description of the functionality of the code, focusing on implementation details and specific behavior.

\end{minipage}
};
\node[below=0.05cm of precise]{\footnotesize (b) Precise};

\end{tikzpicture}
    \end{adjustbox}
        \vspace{-1em}
    \caption{Excerpts from the prompt template used to generate function intents at different precision levels from a Python method implementation.}
    \label{fig:prompt}
            \vspace{-1em}
\end{figure}
Accounting for the prompt design impact on our approach results, in particular, the $M$ transformation from programming to a natural language,
we employ two types of prompts (as shown in Figure~\ref{fig:prompt}):


\inlineheadingit{Broad intent generation (Figure~\ref{fig:prompt}a):} which consists of instructing the LLM to generate a broad (high-level) description of the purpose of the program. This often excludes algorithmic details, making the description ambiguous and incomplete. As a result, the backward transformation $M^{-1}$ has to fill-in the missing details, leading to implementations that are widely different than the original program.

\inlineheadingit{Precise intent generation (Figure~\ref{fig:prompt}b):} which consists of further instructing the LLM to generate a more precise description of the program intent. This description includes important details about the implementation, e.g. steps to perform the computation, algorithms used, or handling of edge cases. As a result, the backward model can produce implementations that are significantly closer to the original program with only small implementation differences. 
Being more precise and preserving of the original program intent, we expect these descriptions to form a better starting point for intent-based mutations, enabling the generation of mutants that originate mainly from the propagation of intent-modifications. 

\section{Evaluation}
%
Our evaluation study aims at answering these questions:
\begin{description}
    \item[\emph{RQ1 (Mutants validity)}] \emph{Can round-trip mutation testing generate valid mutants?}
    \item[\emph{RQ2 (Mutants Diversity)}] \emph{How diverse are round-trip mutations compared to traditional pattern-based mutations?}
     \item[\emph{RQ3 (Test Guidance)}] \emph{How effective and cost-efficient are round-trip mutants in guiding testing towards higher fault detection with fewer tests?}
\end{description}
%

\begin{table}[t]
\vspace{-3em}
\centering
\renewcommand{\arraystretch}{0.8}
\setlength{\tabcolsep}{10pt}
\caption{Summary of buggy programs used in the study}
\vspace{-0.5em}
\label{tab:subject_projects}

\begin{tabular}{lcccc}
\toprule
\textbf{Project} & \textbf{\#Buggy version} & \textbf{\#Buggy Method} \\
\midrule
youtube-dl  & 14  &  15    \\
tqdm  & 3 & 4    \\
black  & 7 & 14    \\
httpie  & 2 & 2   \\
thefuck  & 4 &  5   \\
\bottomrule
\end{tabular}
\vspace{-1em}
\end{table}

\subsection{Experimental setup}
We perform our evaluation on 40 real buggy methods from 30 programs collected from 5 different open-source Python projects, available in the BugsInPy~\cite{widyasari2020bugsinpy} dataset. 
Each fault is represented by a faulty version of the program and a bug fix that resolves the fault. 
We focus on faults that are fixed by changing at least one line of code in a given method. 
We report the number of faults selected per project in Table \ref{tab:subject_projects}.

\inlineheadingit{Mutation Generation} To answer our research questions, we evaluate and compare four variants of round-trip mutation testing: (1) Broad, (2) Precise, (3) $\mu$Broad, and (4) $\mu$Precise. 
The configurations Broad and Precise apply round-trip mutation with the respective prompts directly {\em without} intent mutation (as in Equation~\ref{eq:directgen}). $\mu$Broad and $\mu$Precise map the method to a natural language description, mutate the description, and then map the mutated description back to a mutant. Mutants are generated on the {\em buggy} version of the code and we generate 10 mutants per method using GPT-4o-mini as LLM. 
In addition, we reimplemented\footnote{We originally planned to use MutPy with BugsInPy, but found that MutPy could not be applied due to dependency issues.} the mutation operators defined in MutPy~\cite{mutpy} and use it as a representative baseline for traditional mutation testing, which we note as traditional mutants (trad. mut.).

\begin{figure}
    \vspace{-3em}
    \centering
    \includegraphics[width=\columnwidth]{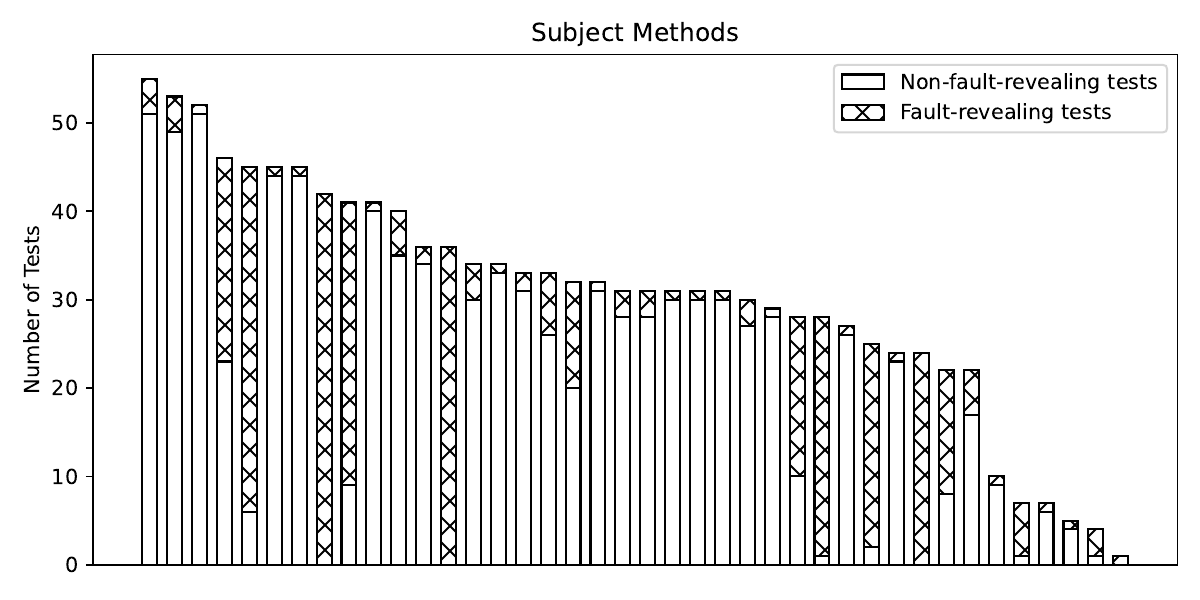}
    \vspace{-2.5em}
    \caption{Number of tests and fault-revealing tests.}
    \label{fig:tests_per_method}
    \vspace{-1em}
\end{figure}

\inlineheadingit{Test Augmentation} We augment the developer test suite with LLM-generated tests. We use GPT-4o-mini to generate tests from the fixed version of each buggy method. In the process, we iteratively refine the generated tests to generate valid tests with high coverage. We ensure that each test suite has at least one fault-revealing test. Overall, with the generated tests, we achieve coverage score of 89.9\% to 94.4\% per method under test. Figure \ref{fig:tests_per_method} shows the total number of tests and the ratio of bug-revealing tests per buggy method.

\inlineheadingit{Mutation Testing Simulation} In our experiments, we simulate a developer using mutation testing to write fault-revealing tests. 
Hence, we generate mutants for the faulty version of the method under test (similar to \cite{ChekamPTH17}). A mutant is killed if there is a measurable difference between the test outcome of the mutated version and the original (faulty) version of the project. Meaning, their executions end with at least one different test result (pass/fail), or at least one assertion receiving a different value, e.g. the returned value from the method under test. All other mutants are surviving, and the developer would write tests to kill them. 
In our simulation, we start with an empty test suite (ts) and incrementally add tests from the full test suite (TS). To this end, we randomly pick a mutant that is not yet killed by the current test suite (ts). Then, we randomly select a killing test (one at a time) from the full test suite (TS) to be added to our current tests (ts). If a mutant cannot be killed, we move on without adding a test and randomly select the next mutant. In this process, a fault is detected as soon as we add a fault-revealing test. Since a fault-revealing test might be selected by accident, we repeat our simulation 100 times and report the average fault detection rate of the resulting test suite at different testing budgets.

\section{Results}
\subsection{RQ1: Can round-trip mutation generate valid mutants?}
To answer \textbf{RQ1}, we start by executing the mutants against the full test suite. We say that a mutant of a Python method is valid iff (1) the mutant passes syntactic validation of the Python interpreter and (2) the mutant is runnable, i.e. it can be tested without raising any runtime errors. 

\inlineheadingbf{Results} Table~\ref{tab:mutants_stats} summarizes our results, reporting the percentages of mutants per approach that are (1) syntactically invalid, (2) surviving, (3) killable but incompetent (i.e. mutants that fail to run due to runtime errors) and (4) killable and runnable. 
As expected, Precise RTM produces the highest number of surviving mutants. The mutants are often closer to the original implementation than the other RTM variants which may result into equivalent mutants. Still, 65.5\% are valid mutants that can be killed by the test suite. Both mutation and instructing the LLM to generate broader descriptions of the intent can help reduce the number of surviving mutants. Yet, the broader descriptions yield the highest number of valid mutants. Although traditional mutation produces a higher ratio of valid mutants, the percentage of killable mutants remains comparable of around $70\%$. 
Whereas traditional mutation produces more surviving mutants ($17.9\%$; nearly $16\%$ more suriving mutants than RTM), round-trip mutation produces relatively more incompetent mutants.

\begin{table}[t]
\vspace{-3em}
\centering
\small
\setlength{\tabcolsep}{1.5pt}
\renewcommand{\arraystretch}{0.9}
\caption{Round-Trip Mutants Vs Traditional Mutants}
\vspace{-0.5em}
\label{tab:mutants_stats}

\begin{tabular}{p{2 cm} p{1.5cm} p{1.5cm} c ll}
\toprule
\textbf{Approach} & 
\textbf{Invalid} & 
\textbf{Surviving} && 
\multicolumn{2}{c}{\textbf{ Killable}} \\
\cmidrule(lr){4-6}
 & & && \textbf{Incompetent} ~~& \textbf{Runnable} \\
\midrule
\textbf{Round-trip } & & & & \\ 
\quad Broad & 0.25\% & 0.25\% && 28.74\% & 70.75\%  \\
\quad Precise & 0.00\% & 2.00\% && 32.75\% & 65.50\%   \\
\quad $\mu\text{Broad}$ & 0.50\% & 0.00\% &&  33.00\% & 66.50\% \\
\quad $\mu\text{Precise}$ & 0.25\% & 1.50\% && 47.75\% & 50.50\%  \\
\midrule
\textbf{trad. mut.} & 0.00\% & 17.93\% && 10.05\% & 72.01\%  \\ 
\bottomrule
\end{tabular}
\vspace{-1em}
\end{table}

\subsection{RQ2: How syntactically diverse are round-trip mutants?}
To answer this question, we investigate how syntactically close mutants are to the original code, as well as between mutants themselves. Prior to any computation, we remove docstrings to focus only on code. We then compute the BLEU distance (1 - BLEU score) of each mutant–original pair and of each mutant–mutant pair. We report the distributions achieved by each approach including the four RTM variants and the syntactic baseline.

\begin{figure}
    \centering
    \includegraphics[width=0.9\columnwidth]{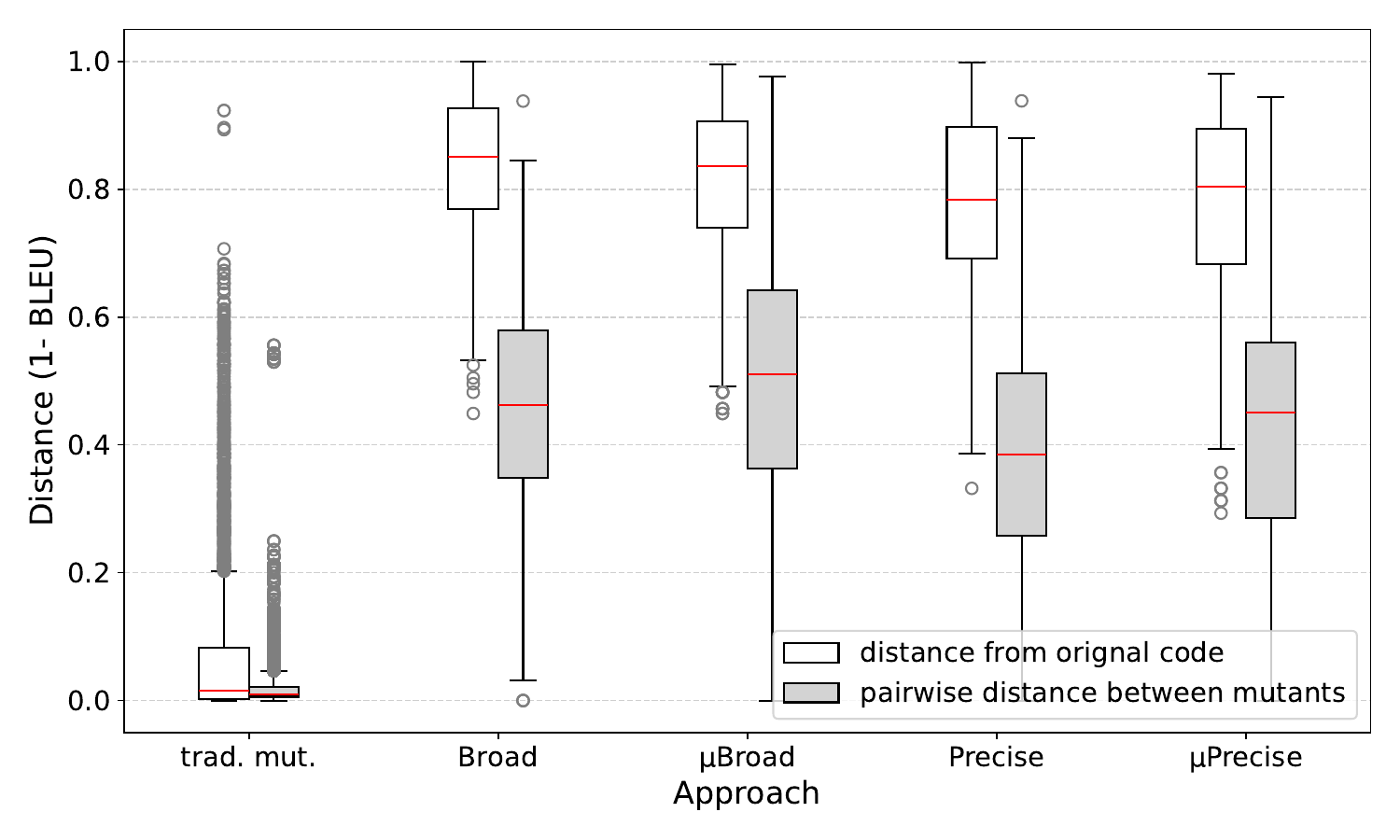}
    \vspace{-1em}
    \caption{Distribution of distances (1 - BLEU score) for code mutations. Boxplots show distances between each mutated method and its original implementation (white), as well as pairwise distances between mutants (grey).}
    \label{fig:distance}
    \vspace{-1em}
\end{figure}

\inlineheadingbf{Results} Figure \ref{fig:distance} depicts the distributions of BLEU distances. Traditional mutants remain very close to both the original implementation and to each other, due to the fine-grained syntactic change. In contrast, RTM variants show a greater syntactic deviation and diversity. 
Mutants generated from Broad intents (which incorporate less information on the code) tend to produce more distant mutants compared to Precise variants, enabling the exploration of a wider spectrum of mutants beyond traditional mutation. Additionally, applying mutation to the intent increases diversity, resulting in mutants that are more varied among themselves. 


\subsection{RQ3: How effective is round-trip mutation testing for test guidance?}
To answer \textbf{RQ3}, we simulate a mutation-guided test-selection scenario, and compare the fault detection ratios of the obtained test-suites by each approach, at each test-selection step. 
In addition, we include a random selection baseline, to contrast the advantage brought by mutation guidance.

\begin{figure}
\vspace{-3em}
    \includegraphics[width=\columnwidth]{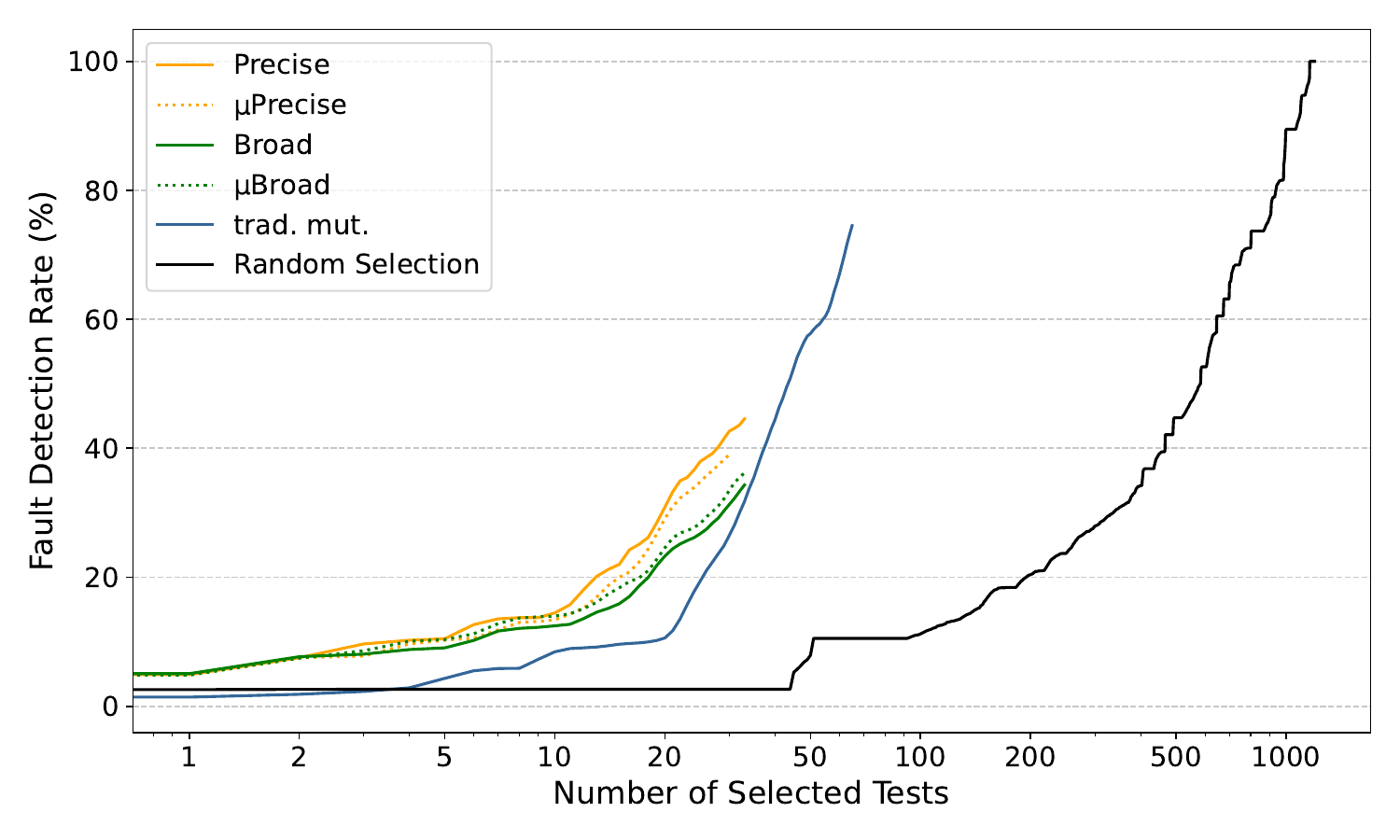}
    \vspace{-2.5em}
    \caption{Mutation testing simulation. The x-axis represents  the number of tests selected by each approach (log scale) and the y-axis represents the fault detection rate of the selected tests across 100 simulations.}
    \label{fig:fd_effort}
    \vspace{-1em}
\end{figure}

\inlineheadingbf{Results} Figure \ref{fig:fd_effort} shows the result of our mutation testing simulation, reporting the percentage of faults that can be detected by selecting $x$ tests (averaged over 100 runs). Although mutation testing achieves a significantly higher fault detection rate than random testing early on, all approaches saturate after \chan{around 66} selected tests. In these cases, all mutants are either killed or are not killable by the full test suite and hence \chan{no further tests are selected.}
Overall, RTM is more effective in selecting fault-revealing tests when the test budget is small (between 1 and 30 tests). 
In fact, it produces 
test suites achieving on average \chan{$\approx$4.3} and \chan{$\approx$1.7}  times higher fault-detection rates than traditional mutants, when selecting only \chan{4} and \chan{30} tests, respectively. 
When selecting about \chan{42 tests} all approaches reach similar fault detection rates, with \chan{RTM} achieving its maximum avg fault detection of \chan{$\approx$44.6\%}, which is slightly higher than the \chan{$\approx$44\%} of traditional mutants.
Traditional mutation testing performs better for larger budgets (beyond \chan{42 tests}), reaching a maximum fault-detection of \chan{$\approx$74.5\%} after the selection of \chan{$\approx$60} tests. 
Perhaps surprisingly, while the precise intent produces less runnable mutants, it achieves a higher fault detection rate in comparison to all other round-trip mutation testing variants.

%

\section{Discussion}
Based on our results, we see round-trip mutation testing as a promising approach for fault detection with a small test budget. However, we identified two current limitations of our approach which we aim to address in the future:

\inlineheadingbf{Impact of Intent Mutation} We initially expected that the mutation of the intent (inline with \cite{hamidi2025intentbasedmutation}) increases fault detection. Yet, our experiments show the contrary: the highest fault detection rate at a given test budget is achieved without applying intent mutation. We suspect that the problem stems from the fact that intent mutation increases the number of incompetent mutants (15\% and 4.26\% increase, respectively) in comparison to original generated intents. Therefore, we are currently investigating two strategies: (1) providing additional context during back translation and (2) using static or dynamic feedback to refine mutants. These strategies are motivated by the observation that functions in real projects are often tightly integrated and have complex dependencies that the code generation LLM might not be aware of.


\inlineheadingbf{Semantic Diversity of RTM Mutants} We further investigate why RTM selects fewer tests than traditional mutation testing in \textbf{RQ3}. To be effective in this scenario, mutants need to show diverse behavior, i.e. breaking different tests. In other words, if all or most mutants are killed by the same tests, the mutation analysis stops when very few tests are selected, leading consequently to an early plateau of test-suite fault detection. 
In fact, when investigating the percentage of semantically {\em unique} mutants (i.e. killed by a different set of tests; see Table \ref{tab:unique_mutants}), we observe that more than \chan{38\%} of traditional mutants are unique, compared to \chan{20\%} of RTM mutants across all its variants. 

\begin{table}[t] 
\vspace{-3em}
\centering
\small
\setlength{\tabcolsep}{1.5pt}
\renewcommand{\arraystretch}{0.9}
\caption{Average Proportion of semantically unique mutants generated by each approach.}
\vspace{-0.5em}
\label{tab:unique_mutants}

\begin{tabular}{p{2 cm} p{2.5cm} p{2.5cm}}
\toprule
\textbf{Approach} & 
\textbf{Runnable} & 
\textbf{All}  \\
\midrule
\textbf{Round-trip } & & \\ 
\quad Broad & 31.90\% & 21.10\%  \\
\quad Precise & 24.30\% & 16.30\%  \\
\quad $\mu\text{Broad}$ & 33.30\% & 22.60\% \\ 
\quad $\mu\text{Precise}$ & 40.90\% & 19.80\% \\
\midrule
\textbf{trad. mut.} & 45.75\% & 38.44\% \\ 
\bottomrule 
\end{tabular}
\vspace{-1em}
\end{table}


\inlineheadingbf{Threats to validity} 
Our results and conclusions might be affected by the nondeterministic nature of the employed LLMs. While it may produce different mutants, generating multiple implementations per program helps mitigate this threat and ensure the reproducibility of the overall results. 
Similar threats could emerge from the random selection of mutants and tests in our simulation. To mitigate this, we run it 100 times for every program and approach.
In addition, relying primarily on automatically-generated tests can threaten the generalizability of our claims. We address this threat by ensuring high coverage, keeping only test suites containing fault-revealing tests and augmenting missing ones with developer-written tests from the dataset. 
A potential threat could arise from the used dataset (the selected faulty programs), and the focus on Python. 
Although we cannot ensure that the results generalize to different projects in other languages with different typing and runtime behaviors, we selected real programs from different open-source projects.
Finally, while our syntax-based mutation baseline is a re-implementation, it uses the same operators as the standard mutation tool for Python, MutPy.

\section{Conclusion}
In this paper, we introduce a novel LLM-based mutation testing approach, which generates mutants by applying a round-trip translation to the input program; from code to intent and back to code. 
By mutating the generated intent, the approach achieves intent-based mutation without relying on a pre-existing intent (e.g. developer-written documentation). 
The evaluation results on real Python bugs show that this approach produces valid mutants and can guide testing toward considerable fault detection rates with only few tests.
Not involving any operators and not requiring any specific knowledge on the programming language at hand, we believe that the approach is easy to use in practice, particularly with the omnipresence of LLMs in all development environments and processes.

\section*{Acknowledgment}
This work is supported by the Luxembourg National Research Fund (FNR) through the CORE project C23/IS/18182513/MiCE.

\bibliographystyle{IEEEtran}
\IEEEtriggeratref{12}
\bibliography{sample}

\end{document}